\pdfoutput=1

\documentclass[aps,twocolumn,superscriptaddress,floatfix,preprintnumbers,nofootinbib]{revtex4-2}

\usepackage{amsmath, float, graphicx,amssymb,multirow,pgfplots,pgfplotstable}
\usepackage{tikz,hyperref}
\usepackage{ulem}
\usepackage{comment}
\usepackage{url}
\usepackage{graphicx}
\usepackage{lipsum}

\usetikzlibrary{shapes.geometric,arrows}

\begin{document}

\begin{flushright}
\end{flushright}

\title{Superheavy Supersymmetric Dark Matter  \\
for the origin of KM3NeT Ultra-High Energy signal
}

\author{Yongsoo Jho}
\affiliation{Department of Physics and IPAP, Yonsei University, Seoul 03722, Republic of Korea}
\author{Seong Chan Park}
\email{sc.park@yonsei.ac.kr}
\affiliation{Department of Physics and IPAP, Yonsei University, Seoul 03722, Republic of Korea}
\affiliation{School of Physics, Korea Institute for Advanced Study, Seoul, 02455, Republic of Korea}

\author{Chang Sub Shin}
\email{csshin@cnu.ac.kr}
\affiliation{School of Physics, Korea Institute for Advanced Study, Seoul, 02455, Republic of Korea}

\affiliation{Particle Theory and Cosmology Group, Center for Theoretical Physics of the Universe,
Institute for Basic Science (IBS), Daejeon, 34126, Korea}
\affiliation{Department of Physics and Institute of Quantum Systems, Chungnam National University, Daejeon 34134, Korea}

\begin{abstract}
We propose an explanation for the recently reported ultra-high-energy neutrino signal at KM3NeT, which shows no clear association with known astrophysical sources. While decaying dark matter in the Galactic Center is a natural candidate, the observed arrival direction strongly suggests an extragalactic origin. We introduce a multicomponent dark matter scenario in which the components are part of a supermultiplet, with supersymmetry ensuring a nearly degenerate mass spectrum among the fields with different spins. In this setup,  a cosmologically long-lived fermionic state decays into a slightly lighter bosonic dark matter state, producing a boosted neutrino spectrum with energy $E_\nu \sim 100$ PeV, determined by the mass difference. The heavy-to-light decay occurs at a cosmological redshift of $z \sim \text{a few}$ or higher, leading to an isotropic directional distribution of the signal.
\end{abstract}

\maketitle


\newcommand{\PRE}[1]{{#1}} 
\newcommand{\ul}{\underline}
\newcommand{\del}{\partial}
\newcommand{\nbox}{{\,\lower0.9pt\vbox{\hrule \hbox{\vrule height 0.2 cm
\hskip 0.2 cm \vrule height 0.2 cm}\hrule}\,}}

\newcommand{\postscript}[2]{\setlength{\epsfxsize}{#2\hsize}
   \centerline{\epsfbox{#1}}}
\newcommand{\gweak}{g_{\text{weak}}}
\newcommand{\mweak}{m_{\text{weak}}}
\newcommand{\mplanck}{M_{\text{Pl}}}
\newcommand{\mstar}{M_{*}}
\newcommand{\sigmaan}{\sigma_{\text{an}}}
\newcommand{\sigmatot}{\sigma_{\text{tot}}}
\newcommand{\sigmaSI}{\sigma_{\rm SI}}
\newcommand{\sigmaSD}{\sigma_{\rm SD}}
\newcommand{\OmegaM}{\Omega_{\text{M}}}
\newcommand{\OmegaDM}{\Omega_{\text{DM}}}
\newcommand{\ipb}{\text{pb}^{-1}}
\newcommand{\ifb}{\text{fb}^{-1}}
\newcommand{\iab}{\text{ab}^{-1}}
\newcommand{\ev}{\text{eV}}
\newcommand{\kev}{\text{keV}}
\newcommand{\mev}{\text{MeV}}
\newcommand{\gev}{\text{GeV}}
\newcommand{\tev}{\text{TeV}}
\newcommand{\pb}{\text{pb}}
\newcommand{\mb}{\text{mb}}
\newcommand{\cm}{\text{cm}}
\newcommand{\m}{\text{m}}
\newcommand{\km}{\text{km}}
\newcommand{\kg}{\text{kg}}
\newcommand{\g}{\text{g}}
\newcommand{\s}{\text{s}}
\newcommand{\yr}{\text{yr}}
\newcommand{\Mpc}{\text{Mpc}}
\newcommand{\etal}{{\em et al.}}
\newcommand{\eg}{{\em e.g.}}
\newcommand{\ie}{{\em i.e.}}
\newcommand{\ibid}{{\em ibid.}}
\newcommand{\Eqref}[1]{Equation~(\ref{#1})}
\newcommand{\secref}[1]{Sec.~\ref{sec:#1}}
\newcommand{\secsref}[2]{Secs.~\ref{sec:#1} and \ref{sec:#2}}
\newcommand{\Secref}[1]{Section~\ref{sec:#1}}
\newcommand{\appref}[1]{App.~\ref{sec:#1}}
\newcommand{\figref}[1]{Fig.~\ref{fig:#1}}
\newcommand{\figsref}[2]{Figs.~\ref{fig:#1} and \ref{fig:#2}}
\newcommand{\Figref}[1]{Figure~\ref{fig:#1}}
\newcommand{\tableref}[1]{Table~\ref{table:#1}}
\newcommand{\tablesref}[2]{Tables~\ref{table:#1} and \ref{table:#2}}
\newcommand{\Dsle}[1]{\slash\hskip -0.28 cm #1}
\newcommand{\met}{{\Dsle E_T}}
\newcommand{\mpt}{\not{\! p_T}}
\newcommand{\Dslp}[1]{\slash\hskip -0.23 cm #1}
\newcommand{\Dsl}[1]{\slash\hskip -0.20 cm #1}

\newcommand{\mB}{m_{B^1}}
\newcommand{\mq}{m_{q^1}}
\newcommand{\mf}{m_{f^1}}
\newcommand{\mKK}{m_{KK}}
\newcommand{\WIMP}{\text{WIMP}}
\newcommand{\SWIMP}{\text{SWIMP}}
\newcommand{\NLSP}{\text{NLSP}}
\newcommand{\LSP}{\text{LSP}}
\newcommand{\mWIMP}{m_{\WIMP}}
\newcommand{\mSWIMP}{m_{\SWIMP}}
\newcommand{\mNLSP}{m_{\NLSP}}
\newcommand{\mchi}{m_{\chi}}
\newcommand{\mgravitino}{m_{\gravitino}}
\newcommand{\mmed}{M_{\text{med}}}
\newcommand{\gravitino}{\tilde{G}}
\newcommand{\Bino}{\tilde{B}}
\newcommand{\photino}{\tilde{\gamma}}
\newcommand{\stau}{\tilde{\tau}}
\newcommand{\slepton}{\tilde{l}}
\newcommand{\snu}{\tilde{\nu}}
\newcommand{\squark}{\tilde{q}}
\newcommand{\mgaugino}{M_{1/2}}
\newcommand{\epsEM}{\varepsilon_{\text{EM}}}
\newcommand{\mmess}{M_{\text{mess}}}
\newcommand{\lmess}{\Lambda}
\newcommand{\nmess}{N_{\text{m}}}
\newcommand{\signmu}{\text{sign}(\mu)}
\newcommand{\Omegachi}{\Omega_{\chi}}
\newcommand{\lambdafs}{\lambda_{\text{FS}}}
\newcommand{\be}{\begin{equation}}
\newcommand{\ee}{\end{equation}}
\newcommand{\bea}{\begin{eqnarray}}
\newcommand{\eea}{\end{eqnarray}}
\newcommand{\beq}{\begin{equation}}
\newcommand{\eeq}{\end{equation}}
\newcommand{\beqn}{\begin{eqnarray}}
\newcommand{\eeqn}{\end{eqnarray}}
\newcommand{\baln}{\begin{align}}
\newcommand{\ealn}{\end{align}}
\newcommand{\lsim}{\lower.7ex\hbox{$\;\stackrel{\textstyle<}{\sim}\;$}}
\newcommand{\gsim}{\lower.7ex\hbox{$\;\stackrel{\textstyle>}{\sim}\;$}}

\newcommand{\ssection}[1]{{\em #1.\ }}
\newcommand{\rem}[1]{\textbf{#1}}

\def\ie{{\it i.e.}\/}
\def\eg{{\it e.g.}\/}
\def\etc{{\it etc}.\/}
\def\calN{{\cal N}}

\def\mptwo{{m_{\pi^0}^2}}
\def\mp{{m_{\pi^0}}}
\def\sqtsn{\sqrt{s_n}}
\def\sqtsn{\sqrt{s_n}}
\def\sqtsn{\sqrt{s_n}}
\def\sqts0{\sqrt{s_0}}
\def\Dsqts{\Delta(\sqrt{s})}
\def\Omegatot{\Omega_{\mathrm{tot}}}

\newcommand{\changed}[2]{{\protect\color{red}\sout{#1}}{\protect\color{blue}\uwave{#2}}}


\section{Introduction}

The KM3NeT collaboration has recently reported their investigation concerning the detection of an Ultra-High-Energy neutrino event, designated as KM3-230213A \cite{KM3NeT:2025npi}. The neutrino energy for this event is estimated to be $E_\nu = 220_{-146}^{+2380}$ PeV with a 90\% confidence interval. This represents the most energetic neutrino observed to date. Multiple hypotheses have been proposed regarding the source of KM3-230213A \cite{Shimoda:2024qzw, Li:2025tqf, Muzio:2025gbr, KM3NeT:2025vut, Dzhatdoev:2025sdi, Podlesnyi:2025aqb, Neronov:2025jfj, DeLaTorreLuque:2025zsv, Boccia:2025hpm, Brdar:2025azm, Borah:2025igh, Kohri:2025bsn, Narita:2025udw, Boccia:2025hpm, Simeon:2025gxd, Jiang:2025blz, Alves:2025xul, Wang:2025lgn, Barman:2025bir}. Notably, no definitive associations with particular astrophysical point origins within galactic and extragalactic regions have been established so far \cite{KM3NeT:2025aps, KM3NeT:2025bxl}.

We propose the decay of Superheavy Dark Matter as a potential explanation for the observed signal at KM3NeT. Specifically, we focus on the distinctive characteristics of the signal:
\begin{itemize}
    \item The neutrino energy: $E_\nu \sim O(100)\, \text{PeV}$
    \item The signal originates from a direction nearly opposite to the Galactic Center.
\end{itemize}
Clearly, the second feature requires an explanation for the absence of a signal from the Galactic Center, which would naturally be expected in a conventional dark matter interpretation~\cite{Park:2015gdo, Park:2012xq}.

To address these challenges, we propose a multicomponent dark matter scenario with a heavy-to-light mass degeneracy, $\Delta M / M \ll 1$. Notably, this framework naturally emerges in a supersymmetric (SUSY) setting \cite{susy1:KaneShifman,Ellis:1983ew,Nilles:1982dy}, where mass degeneracy is an intrinsic feature of supersymmetry. Moreover, when the SUSY-breaking scale is significantly smaller than the mass term, small mass splittings remain radiatively stable \cite{Witten:1981nf,Dimopoulos:1981zb,Susskind:1982mw,Girardello:1981wz,Hall:1990ac}. In this scenario, the heavier component decays into the lighter one, producing a boosted neutrino spectrum with energy $E_\nu \sim \Delta M \ll M$.

The near-degeneracy ensures that the dark matter produced from these decays remains non-relativistic, preserving its role as cold dark matter. Furthermore, the lifetime of the heavier component can be easily controlled, allowing it to be safely set shorter than the age of the Universe. This guarantees that the resulting neutrino spectrum originates entirely from extragalactic sources. Given these characteristics, we refer to this scenario as \textit{superheavy supersymmetric dark matter}.

This letter is organized as follows: In Section~\ref{sec:conventional_decaying_DM}, we discuss the conventional decaying dark matter (DM) scenario, where the expected signal from the Galactic Center is typically $5$–$10$ times stronger than that from the opposite direction. In Section~\ref{sec:model}, we present our two-component superheavy dark matter model and explain how it accounts for the KM3NeT signal. In Section~\ref{sec:signal}, we analyze the expected signatures in neutrinos and associated gamma-rays, considering current observational constraints. Finally, in Section~\ref{sec:conclusion}, we summarize our scenario for the KM3NeT ultra-high-energy (UHE) event and discuss its implications for UHE cosmic rays (UHECRs), UHE neutrinos (UHE$\nu$), and future research directions.

\section{Conventional Dark Matter scenarios \label{sec:conventional_decaying_DM}}

\begin{figure}[h]
\centering
{\includegraphics[width=0.44\textwidth]{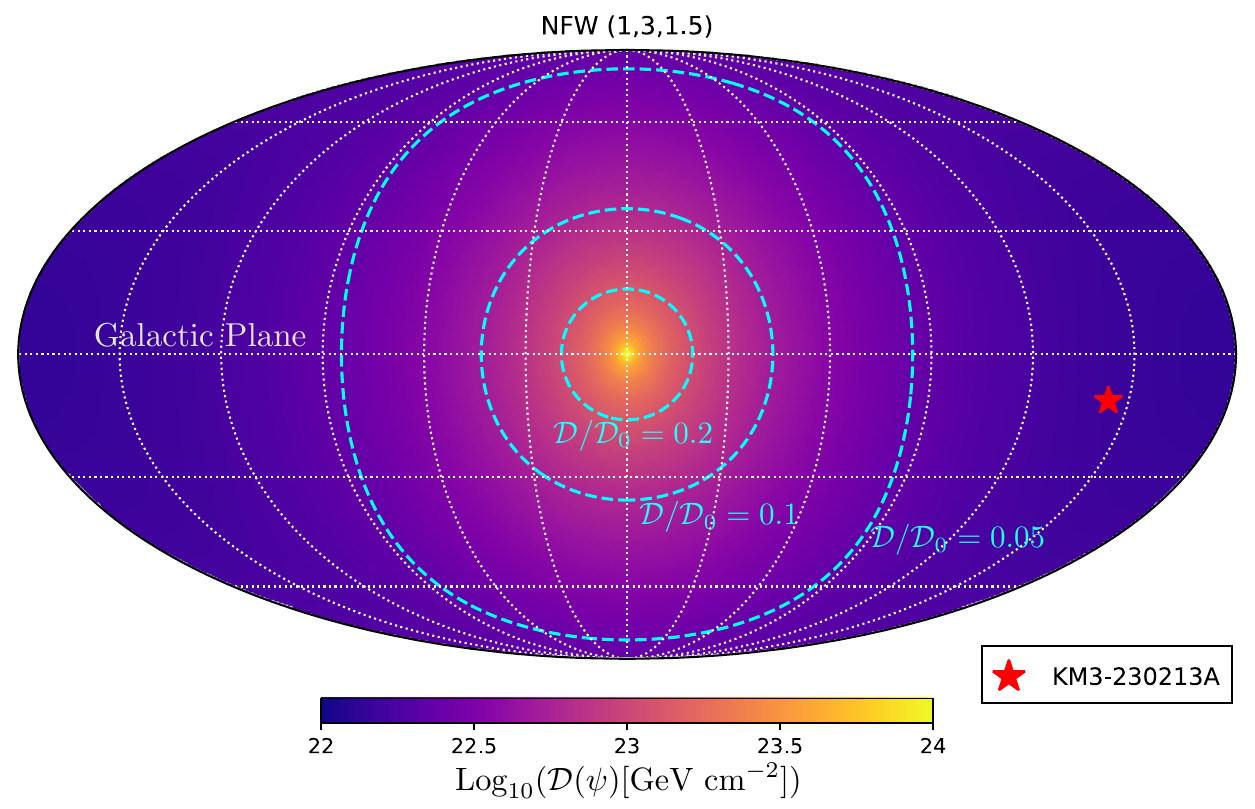}}
{\includegraphics[width=0.48\textwidth]{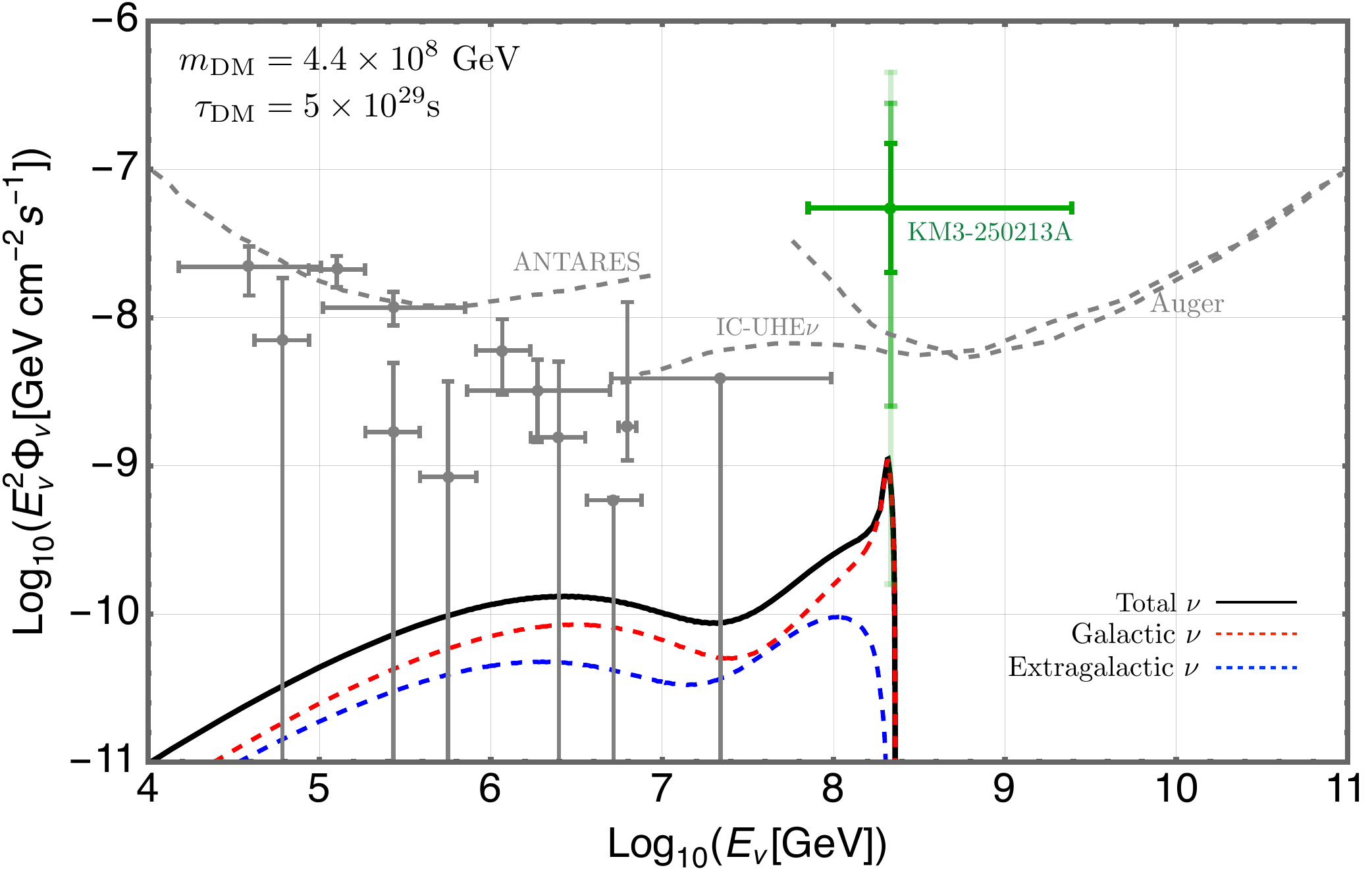}}
\caption{The angular distribution of galactic component of neutrino flux from conventional decaying Dark Matter using NFW(1,3,1.5) profile (Top) and the neutrino flux spectrum (Bottom). In the top panel, we show the $\mathcal{D}$-factor $\mathcal{D}(\psi) = \int \rho(l,\psi) \ dl$ on the Galatic coordinate map. Cyan dashed lines correspond to $\mathcal{D}(\psi)/\mathcal{D}(\psi_0) = 0.2,0.1,0.05$ from inner to outer contours where $\psi$ is the angle between Galactic Center and the direction of line-of-sight. Here we consider the cutoff angle $\psi_0 = 1^\circ$ for the normalization. We overlap the direction of KM3-250213A neutrino origin as a red-star marker whose position is $(l,b)=(216.06^\circ,-11.13^\circ)$ in the Galactic coordinates.  In the bottom panel, we show galactic (dashed red) and extragalactic (dashed blue) contribution to neutrino flux, and observational constraints on the diffuse flux from IceCube \cite{IceCube:2020wum, Abbasi:2021qfz, IceCube:2021rpz, IceCube:2018fhm}, Auger \cite{PierreAuger:2023pjg} and ANTARES \cite{ANTARES:2024ihw} (gray lines).}
\label{fig:Conventional_decaying_DM}
\end{figure}

In this section, we discuss the conventional decaying dark matter scenario. As we see in top panel of Figure.~\ref{fig:Conventional_decaying_DM}, the angular distribution of Galactic component of neutrinos from decaying DM is concentrated around the Galactic Center. Since the angle between the KM3NeT UHE$\nu$'s direction (RA $94.3^\circ$, Dec $-7.8^\circ$) and the Galactic Center (RA $266.4^\circ$, Dec $-29.0^\circ$) is $\psi_{\rm KM3NeT} = 142.5^\circ$, it is natural to see more energetic neutrinos at the Galactic Center. Even including extragalactic contribution, this anisotropy is not dramatically relaxed. For instance, let us define a ratio between energy-integrated fluxes for different two directions as
\begin{eqnarray}
\mathcal{P} \equiv \frac{ \int_{E_{\min}}^{E_{\max}} dE \frac{d^2 \Phi}{dE d\Omega} \Bigr |_{\rm GC} }{ \int_{E_{\min}}^{E_{\max}} dE \frac{d^2 \Phi}{dE d\Omega} \Bigr |_{\rm KM3} }
\end{eqnarray}
We read the ratio $\mathcal{P}$ for three different DM profiles (NFW(1,3,1.5), NFW(1,3,1) and Isothermal) and two different energy windows ([72, 2600] PeV (90CL) and [170, 270] PeV (Narrow)) as
\begin{eqnarray}
&& \mathcal{P}_{\rm NFW1315}^{\rm Narrow} = 79.6,  \quad \quad \mathcal{P}_{\rm NFW1315}^{\rm 90CL} = 75.1, \nonumber \\
&& \mathcal{P}_{\rm NFW131}^{\rm Narrow} = 12.8, \quad \quad \, \, \mathcal{P}_{\rm NFW131}^{\rm 90CL} = 9.0, \nonumber \\
&& \mathcal{P}_{\rm Iso}^{\rm Narrow} = 7.9, \quad  \quad \quad \, \mathcal{P}_{\rm Iso}^{\rm 90CL} = 5.3, \nonumber
\end{eqnarray}
We use the cutoff angle $\psi_0 = 1^\circ$ to estimate GC flux for cuspy profiles. The parametrization NFW($\alpha$,$\beta$,$\gamma$) is defined as 
\bea \rho(r) = \rho_0 \cdot \left(\frac{r_s}{r}\right)^{\gamma} \Bigl(1+\Bigl(\frac{r}{r_s}\Bigr)^\alpha \Bigr)^{\frac{\gamma-\beta}{\alpha}}.\eea
As we see in bottom panel of Figure.~\ref{fig:Conventional_decaying_DM}, the spectral shape of flux is more narrow in the galatic contribution compared to the extragalactic flux, due to the redshift effect.   
In any cases, one expects the expected different number of events with a factor of $5-80$ around $E_\nu \sim \mathcal{O}(100)$ PeV in the conventional decaying DM scenario. Nevertheless, since we have only one event within the $\mathcal{O}(100)$ PeV energy window, it obviously requires more enhanced statistics based on observational data set with a larger effective area experiments. For now, let us keep this possibility, and proceed to construct a new scenario which gives the isotropic and extragalactic-dominant signals at high energies.

\section{Model \label{sec:model}}

\begin{figure}
{\includegraphics[width=0.35\textwidth]{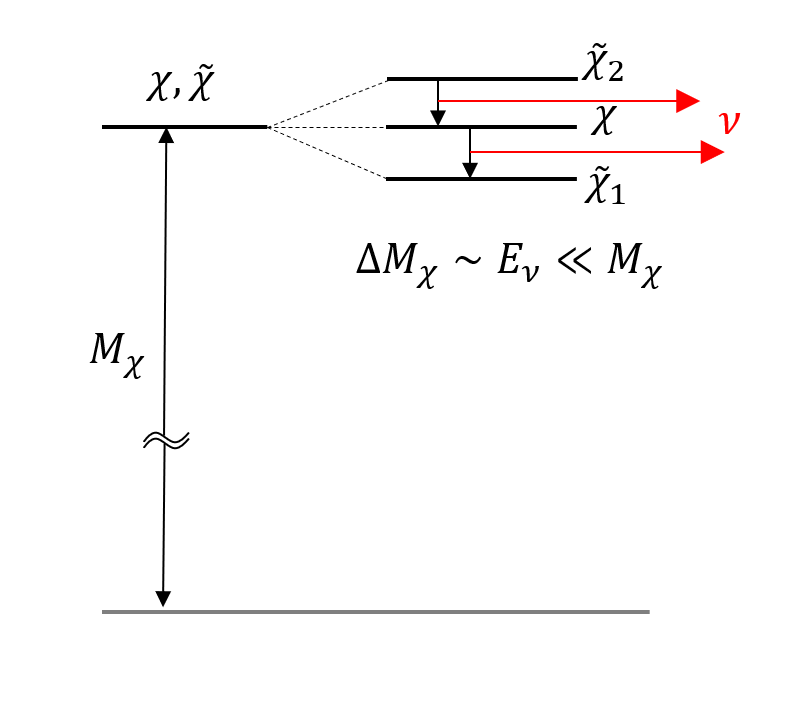}}
\caption{Scheme of a model for superheavy supersymmetric dark matter. The dark matter supermultiplet possesses a large supersymmetric mass term, \( M_\chi \), with a small mass splitting induced by SUSY-breaking effects. This small mass splitting remains radiatively stable compared to the dark matter mass.  
Each component field contributes equally to the dark matter density through their supersymmetric production mechanism. The long-lived heavier component of the supermultiplet decays into the lightest one, emitting a highly energetic neutrino (directly, secondarily, or via oscillation) with energy \( E_\nu \sim \Delta M_\chi \ll M_\chi \) at production time. Consequently, any late-time decay of \( \chi \) and \( \tilde{\chi}_2 \) remains consistent with cosmological constraints on dark matter density.
} \label{fig:Scheme_susyDM}
\end{figure}

\subsection{Dark matter mass spectrum}

The basic scheme of our model is shown in Fig.~\ref{fig:Scheme_susyDM}. We consider a dark matter supermultiplet, whose dominant mass term is given by the supersymmetric mass $M$. We allow the possibility that $M$ can be as large as a grand unified scale, i.e., $10^{16}\,{\rm GeV}$. Since the SUSY breaking scale in our set-up is much smaller than $M$, it generically predict highly degenerated mass spectrum among dark matter components.

Dark matter is given as the component field of the SM singlet chiral superfield $X$ \cite{Salam:1974yz},
\begin{equation}
X = \tilde\chi + \sqrt{2} \theta \chi + \theta^2 F_\chi.
\end{equation}
Here, $\tilde\chi$ is a complex scalar, $\chi$ is the its fermionic superpartner, and $F_X$ is the auxiliary scalar field.

Since we consider superheavy dark matter with a mass much greater than the SUSY breaking scale, it is natural that the dominant mass contribution arises from the superpotential
\begin{equation}
W = \frac{1}{2} {\cal M} X^2,
\end{equation}
where the spurion superfield ${\cal M}$ is given by
\begin{equation}
{\cal M} = M + \theta^2 F.
\end{equation}
The scalar component of ${\cal M}$ provides a supersymmetric mass $M$ for both the scalar and fermionic components, while the $F$-term contribution induces mass splitting among the scalars.

The spurion method efficiently captures the SUSY breaking effects in the mass spectrum of the supermultiplet as long as $M^2 \gg F$. Superheavy supersymmetric dark matter is realized for sufficiently large $M$. The quadratic potential for the scalar and fermion components is obtained as
\begin{equation}
V =  \frac{1}{2} M (\chi \chi + h.c.)  + M^2 |\tilde\chi|^2 + \frac{1}{2} F  (\tilde\chi^2 + h.c.).
\end{equation}
The resulting mass splitting between the scalar and fermionic components is given by
\begin{equation}
M_{\tilde\chi_{1,2}} = M \pm \frac{F}{M}, \quad M_\chi = M.
\end{equation}
Thus, the relative mass difference is
\begin{equation}
\frac{\Delta M_\chi}{M_\chi} \equiv \frac{M_{\chi} - M_{\tilde\chi_1}}{M_\chi} =\frac{M_{\tilde\chi_2} - M_\chi}{M_\chi}=  \frac{F}{M^2} \ll 1.
\end{equation}
This very degenerate mass spectrum remains radiatively stable as long as $F \ll M^2$, since $F$ serves as the order parameter for SUSY breaking, ensuring that all other mass splittings are also proportional to $F$.

The lightest component of $X$ (i.e., $\tilde\chi_1$) can serve as a stable dark matter candidate due to the $\mathbb{Z}_2$ symmetry, $X\to -X$. However, if the interaction between $X$ and other fields is sufficiently suppressed, the heavier states $\chi$ and $\tilde\chi_2$ may also contribute to dark matter due to their long lifetimes.
We can first think of the interaction that determines the dominant decay channel of heavier dark matter.

\subsection{Neutrino flux from the decay of heavier dark matter components}

Although the production mechanism of heavy dark matter in the early universe is not unique~\cite{Carney:2022gse}, it is reasonable to expect that the supersymmetric contribution dominates over the SUSY-breaking one, as dark matter is primarily produced through high-energy interactions exceeding the SUSY-breaking scale.
Therefore, it is natural to assume that the initial abundances of $\tilde\chi_1$, $\chi$, and $\tilde\chi_2$ are approximately equal, i.e., $\rho_{\tilde\chi_{1,2}}\sim \rho_{\chi}$.

The heavier components eventually decay into $\tilde\chi_1$.
Due to the nearly degenerate mass spectrum, the resulting $\tilde\chi_1$ is not boosted and remains a component of cold dark matter.
The key parameters governing the neutrino flux from these decays are the fraction of heavy components in the dark matter density, $f_\chi$, the total decay rate of the heavier component, $\Gamma_{\chi}$, and the decay rate of the heavier component into active neutrinos, given by $\text{Br}_{\nu} \cdot \Gamma_\chi$.

The isotropic extragalactic neutrino flux is then expressed as
\begin{eqnarray} 
\frac{d\Phi_\nu}{dE_\nu} &=& \frac{\text{Br}_{\nu} \Gamma_\chi}{ 4\pi M_\chi} \frac{f_\chi \rho_{\rm DM} c}{H_0} \nonumber\\ 
& & \times \int_0^{z_{\max}} \frac{dz}{1+z}
\frac{e^{- \Gamma_\chi t(z)}}{\sqrt{\Omega_m (1+z)^3 + \Omega_\Lambda}} \frac{dN_\nu}{dE_\nu'} 
\label{eq:Total_nu_flux} \end{eqnarray} 
where $E_\nu'=(1+z)E_\nu$. The upper limit of the redshift integral, $z_{\max}$, can, in principle, extend up to the neutrino decoupling redshift, $z_{\rm dec.} \simeq 10^{10}$, but for the parameter range of interest, the dominant contribution typically arises from $0\leq z \lsim 10$. 
The present-day dark matter density is given by
$\rho_{\rm DM}=1.15 \times 10^{-6}\,\rm GeV {\rm cm}^{-3}$,
and the Hubble radius is $c/H_0=1.37\, \times 10^{28}{\rm cm}$.
The matter and dark energy contributions to the total energy density are
$\Omega_m=0.315$ and $\Omega_\Lambda=0.685$, respectively \cite{Planck:2018vyg}.
Since we consider the case where all heavier components decay before the present time, i.e., $\Gamma_\chi <H_0$, only the extragalactic contribution remains.

The necessary conditions for the branching fraction and mass splitting can be estimated from observations. The observed KM3NeT neutrino flux is given by
\begin{eqnarray} E_\nu^2 \frac{d\Phi_\nu}{dE_\nu} = 5.8^{+10.1}_{-3.7}\times 10^{-8}\left(\frac{\rm GeV}{\cm^2 \sec {\rm sr}}\right) \end{eqnarray} for $E_\nu =(0.072-2.6)\times 10^9{\rm GeV}$ \cite{KM3NeT:2025npi}.
Given the present dark matter density and the relation $E_\nu = \Delta M_\chi/(1+z_\chi)$, where $z_\chi$ is the redshift at $t=1/\Gamma_\chi$, the estimated flux for our model is
\begin{eqnarray} E_\nu^2 \frac{d\Phi_\nu}{dE_\nu} &\sim & \frac{{\rm Br}_\nu f_\chi  }{4\pi M_\chi}\frac{\Delta M_\chi }{ (1+z_\chi)} \rho_{\rm DM} c \nonumber\\ &=& 2.5\times 10^{-8}\left(\frac{\rm GeV}{\cm^2 \sec {\rm sr}}\right)  \nonumber\\ &&\times \left(\frac{{\rm Br}_\nu f_\chi}{10^{-3}}\right)\left(\frac{\Delta M_\chi }{10^{-7} M_\chi}\right)\left(\frac{10}{1+z_\chi}\right). \end{eqnarray}
A degenerate mass spectrum of this kind can naturally arise within supersymmetry, while the decay products and lifetimes depend on the specific model construction. The SUSY breaking mass $F/M \sim 10^8-10^9 {\rm GeV}$ is still much higher than the weak scale. This is consistent with the null result of SUSY searches \cite{Allanach:2024suz,ParticleDataGroup:2022pth,Martin:1997ns}. The gravitino and superpartners of the Standard Model particles are heavy enough that they can decay much earlier than the BBN epoch if the $R$-parity violating terms exist. Therefore only $X$ supermultiplet remains as a dark matter.  A detailed discussion of R-parity violation (RPV) and its implications for BBN and proton stability is provided in Appendix~\ref{app:RPV}.

To allow for variations in the branching fraction into active neutrinos, we assume that the dominant decay channel of the dark matter supermultiplet is invisible.
Introducing a singlet chiral superfield,
\bea S= \tilde\nu_s + \sqrt{2}\theta \nu_s + F_s, \eea
the superpotential governing the decay of the heavier dark matter components is
\bea W_D = \frac{1}{2}\lambda_s S X^2, \eea
where $\lambda_s$ is a free parameter.
For a mass much smaller than $M_\chi$, this term induces the decays
$\chi\to \tilde\chi_1 + \nu_s$ and $\tilde\chi_2 \to \chi + \nu_s$.
Assuming this decay channel dominates, the heavier dark matter decay rate is approximately
\bea \Gamma_{\tilde\chi_2 \to \chi \nu_s} \simeq \Gamma_{\chi \to \tilde\chi_1 \nu_s} \simeq \frac{\lambda_s^2 M_\chi }{8\pi}\left(\frac{\Delta M_\chi}{M_\chi}\right)^2. \eea
Note that the degeneracy provides a strong kinetic suppression for the decay rate.

We consider different possibilities for the decay into visible particles. The first case is that dark matter has a direct interaction with neutrinos. The corresponding minimal superpotential is \begin{equation} \Delta W_{\rm I} = \frac{LH_u X^2} {2\Lambda}. \end{equation} where $\Lambda$ is assumed to be much larger than the Planck scale.
This term leads to the component Lagrangian
\begin{equation} \frac{\nu h \tilde\chi \chi}{\Lambda} + h.c.. \label{eq:eff_op_3body} \end{equation} which allows for three-body dark matter decay via the channels
\begin{equation} \tilde\chi_2 \to \chi + \nu + h,\quad \chi \to \tilde\chi_1 + \nu + h
\end{equation} with the decay rate
\bea \Gamma_{\tilde\chi_2 \to \chi \nu h} \simeq \Gamma_{\chi\to \tilde\chi_1\nu h} \simeq \frac{M_\chi^3}{24 \pi^3 \Lambda^2} \left(\frac{\Delta M_\chi}{M_\chi}\right)^4. \eea
A similar decay channel to neutrinos from long-lived dark matter components has been considered without supersymmetry \cite{Bandyopadhyay:2020qpn}. 

The second case is that the decay into active neutrinos is indirect. The simplest possibility is that the active neutrino is generated from the oscillation of the singlet fermion $\nu_s$ via \begin{equation} \Delta W_{\rm II} = y_s L H_u S. \end{equation}
This leads to the decay $\chi \to \tilde\chi_1 + \nu_s$ and $\tilde\chi_2 \to \chi + \nu_s$, followed by $\nu_s \to \nu$ via oscillation \cite{Dasgupta:2021ies}, with a mixing angle determined by $y_s$ and the masses of neutrinos, assuming the sterile neutrino mass is sufficiently small, on the order of ${\rm eV}$.
This ultimately results in a two-body decay spectrum
\bea \chi\to \tilde\chi_1 + \nu \eea
with a specific branching fraction into neutrinos, ${\rm Br}_\nu$.

\subsection{Production Mechanisms in the Early Universe}

Heavy dark matter with masses in the range $M\sim10^{11}\text{--}10^{16}\,\mathrm{GeV}$ can be produced by several largely model-independent, non-thermal mechanisms. In particular, (i) gravitational particle production at the end of inflation and (ii) Hawking evaporation of light primordial black holes (PBHs) in the early Universe are both efficient over complementary mass intervals. Importantly, because these mechanisms depend primarily on the expansion history and on $M$ rather than on specific dark-sector couplings, nearly degenerate components of a supermultiplet are produced in comparable amounts.\footnote{In supersymmetric setups, superpotential couplings within a multiplet are aligned, and purely gravitational couplings are universal; in addition, our scenario assumes extremely small interactions with SM fields, so thermal production is negligible.}

(i) For a given inflationary Hubble scale $H_{\rm inf}$, when $M\lesssim H_{\rm inf}$ vacuum fluctuations and the non-adiabatic evolution across the end of inflation can generate a relic abundance controlled by the inflationary scale and the reheating history. A convenient parametric estimate is
\begin{equation}
\Omega_{\rm DM} h^2  = \mathcal{C}
\left(\frac{M}{10^{11}\,\mathrm{GeV}}\right)^{2}
\left(\frac{T_{\rm RH}}{10^{9}\,\mathrm{GeV}}\right),
\label{eq:grav-prod}
\end{equation}
where $\mathcal{C}=\mathcal{O}(1)$ encodes the mild dependence on the inflaton potential and the details of reheating (see, e.g., \cite{Chung:2001cb,Kolb:2023ydq}). Gravitational production is particularly efficient for $M\lesssim 10^{14}\,\mathrm{GeV}$.

(ii) Hawking evaporation provides another robust, non-thermal source of superheavy quanta. For heavier dark matter, $M\gtrsim 10^{14\text{--}16}\,\mathrm{GeV}$, evaporation of light PBHs can reproduce the observed dark-matter abundance. Using standard instantaneous-emission results around the evaporation epoch $a_{\rm ev}$, one finds for the comoving number density 
\begin{align}
\hskip -0.25cm n_{\rm DM}(a_{\rm ev}) &\simeq
\left(\frac{1\,\mathrm{g}}{M_{\rm PBH}}\right)^{4} 
\left(\frac{0.1\,M_{\rm Pl}}{M}\right)^{2}  \left(\frac{4\beta}{10^{-24}}\right)\mathrm{GeV}^{3}
\end{align}
and the required initial PBH fraction $\beta\equiv\rho_{\rm PBH}/\rho_{\rm tot}$ (to obtain the CDM abundance today)
\begin{align}
 \beta  &\simeq
3.45\times10^{-20}
\left(\frac{1\,\mathrm{g}}{M_{\rm PBH}}\right)^{1/2}
\left(\frac{M}{10^{16}\,\mathrm{GeV}}\right),
\end{align}
valid for initial PBH masses $M_{\rm PBH}\sim 1\,\mathrm{g}$ to $10^{8}\,\mathrm{g}$ \cite{Carr:2020gox,Cheek:2021odj,Hooper:2019gtx,Masina:2020xhk}. These relations illustrate that the viable $\beta$ scales only with $M_{\rm PBH}$ and $M$, with no strong model dependence in the dark sector. Recent studies have emphasized that this mechanism naturally yields the required dark-matter abundance in the above mass window \cite{Masina:2020xhk,Choi:2025hqt}.

In summary, for the exceedingly small couplings to the SM assumed in our scenario, thermal production is negligible, and non-thermal (semi-)gravitational mechanisms dominate. Because their efficiencies primarily track $M$ and the background cosmology, the nearly degenerate components of the supersymmetric multiplet are produced in similar proportions, consistent with our multicomponent framework.

\section{Signal \label{sec:signal}}

As we notice in the previous section, in our model, there are three components for CDM at high scale $M$ and the spectrum has small mass splittings characterized by a scale $\Delta M_\chi$. We consider two scenarios related to neutrino signals:
\begin{itemize}
\item \textit{Scenario I}: Heavier CDM components decay into lighter CDM, a active neutrino and SM higgs via 3-body decay, $\chi_{\rm heavy} \to \chi_{\rm light} + \nu_\alpha + h$ \, ($\alpha=e,\mu,\tau$).
\item \textit{Scenario II}: Heavier CDM decay into lighter one and a light sterile neutrino via 2-body decay $\chi_{\rm heavy} \to \chi_{\rm light} + \nu_s$, and $\nu_s$ oscillates during its propagation with an active-sterile mixing $\theta_{\alpha s}$.
\end{itemize}
Of course these two scenarios can simultaneously occur and contribute to neutrino signals in general.

\begin{figure*}[!t]
\centering
\includegraphics[width=\textwidth]
{{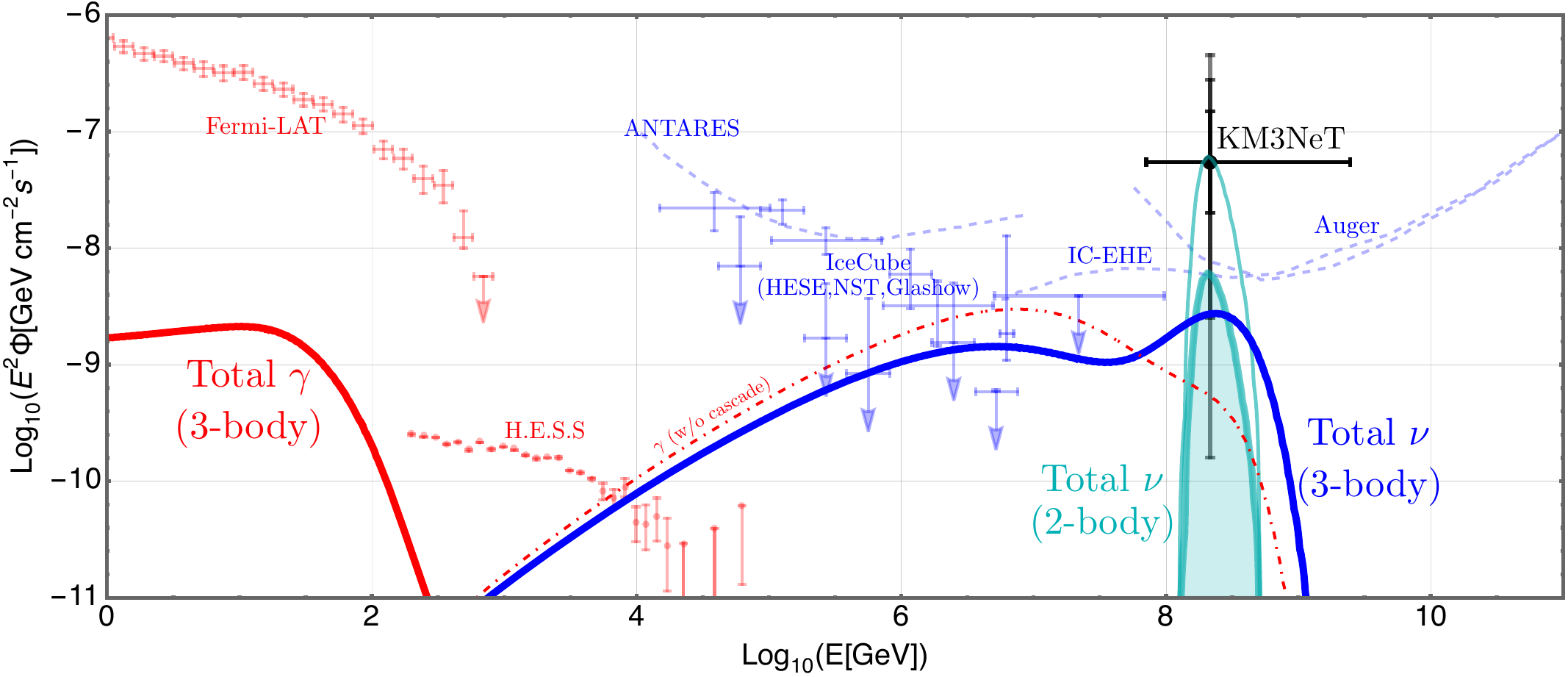}}
\caption{{\bf The landscape of total neutrino and gamma-ray fluxes from Scenario I ($\chi \to \tilde\chi_- + \nu_\alpha + h$) and Scenario II ($\chi \to \tilde\chi_- + \nu_s$ and $\nu_s \to \nu_\mu$ via oscillation)}. We show the total flux of neutrinos and gamma-rays in Scenario I as thick blue and thick red lines respectively. The total flux of neutrinos in Scenario II is shown as thick green line. In order to provide demonstrative examples, we choose the parameters as $M_\chi = 10^{16}$ GeV, $\Delta M_\chi = 3 \times 10^9$ GeV, $f_\chi = 0.6$, $\text{Br}_\nu = 10^{-2.6}$ for Scenario I, and $M_\chi = 6 \times 10^{12}$ GeV, $\Delta M_\chi = 5 \times 10^8$ GeV, $f_\chi =0.6$, $\text{Br}_\nu =10^{-4}$ for Scenario II. The lifetime of heavier component $\chi$ is always assumed to be $\tau_\chi \equiv \Gamma_\chi^{-1} = 1$ Gyr. For the details, see the main text. The upper limits on isotropic diffuse fluxes of gamma-rays (from H.E.S.S \cite{HESS:2016pst} and Femri-LAT \cite{Fermi-LAT:2014ryh}) and neutrinos (from ANTARES \cite{ANTARES:2024ihw}, IC-HESE \cite{IceCube:2020wum}, IC-NST \cite{Abbasi:2021qfz}, IC-Glashow \cite{IceCube:2021rpz}, IC-EHE \cite{IceCube:2018fhm} and Auger \cite{PierreAuger:2023pjg}) are shown as faint red and blue lines, respectively. The desired neutrino flux based on KM3NeT effective area is indicated by black and gray lines \cite{KM3NeT:2025npi}. In Scenario I, gamma-ray flux without attenuation and electromagnetic cascade is shown as a dot-dashed red line for comparison. In Scenario II, we also show the case with $\text{Br}_\nu = 10^{-3}$ as upper thin green line.}
\label{fig:flux_EGdominant_DM_total}
\end{figure*}

In both sceanrios, the processes of decays are $(\chi_{\rm heavy}, \chi_{\rm light}) = (\tilde\chi_2, \chi)$ and $(\chi, \tilde\chi_1)$. These two processes contribute to the neutrino signal similarly. For the case with relatively shorter lifetime compared to the age of our Universe $\tau_\chi (=\Gamma_\chi^{-1}) \lsim \mathcal{O}(1)$ Gyr, because the dominant decaying region is very distant from the Earth ($\gsim \mathcal{O}(1)$ Gpc), we mainly focus on neutrinos and gamma-rays in this work.

{\bf \textit{Neutrinos.}} The mean-free-path of neutrinos in the extragalactic propagation is much larger than $\mathcal{O}(\text{Gpc})$ if the interaction between neutrinos cosmic target particles is dominated by electroweak interactions. The resulting neutrino flux at Earth is obtained by only redshifting the injection spectrum $dN_\nu/dE_\nu$ at each redshift $z$. 

The total neutrino spectrum is basically given by Eq.~(\ref{eq:Total_nu_flux}). The total neutrino spectrum $dN_\nu/dE_\nu$ is given by
\begin{eqnarray}
\frac{dN_\nu}{dE_\nu} & = & \Bigl (\frac{dN_\nu}{dE_\nu} \Bigr )_{\rm prim.} + \Bigl(\frac{dN_\nu}{dE_\nu}\Bigr)_{\rm sec.}
\end{eqnarray}
where $(dN_\nu/dE_\nu)_{\rm prim.}$ is a monochromatic primary spectrum directly produced from hard decay processes ($\tilde\chi \to \chi \nu h$ or $\tilde\chi \to \chi \nu_s$) and $(dN_\nu/dE_\nu)_{\rm sec.}$ is a continuum tail of secondary neutrinos from electroweak/strong cascades, hadronization, and decays of soft mesons. For the evaluation of secondary spectrum, we use HDMSpectra \cite{Bauer:2020jay}.\footnote{ In order to estimate the signal flux in Scenario I, we note that the mean energy of light particles $\langle E_{\nu,h} \rangle \approx \frac{\Delta M}{2}$ for the case with $|\mathcal{M}|^2 = constant$, and $\langle E_{\nu,h} \rangle \approx \frac{2\Delta M}{5}$ for the case with the effective operator (\ref{eq:eff_op_3body}) in the limit $\Delta M \ll M$. In Fig.~\ref{fig:flux_EGdominant_DM_total}, as an approximation we take $E_{\nu,h} = \frac{\Delta M}{2}$ for the energy of neutrino and Higgs boson at the rest frame of the heavy DM particle, to generate the flux of secondary photons and neutrinos from the decay of Higgs.} We have both primary and secondary spectrum in Scenario I ($\chi_h \to \chi_l + \nu + h$), and only primary spectrum in Scenario II ($\chi_h \to \chi_l + \nu_s$).

{\bf \textit{Gamma-rays.}} Compared to neutrinos, the cosmological background is much less transparent to photons. For $\mathcal{O}(\text{PeV})\lsim E_\gamma \lsim \mathcal{O}$(EeV), the mean-free-path is even less than $\mathcal{O}$(1) Mpc \cite{Murase:2010va, Murase:2012xs}. Attenuation and electromagnetic cascade processes significantly affect its propagation \cite{Lee:1996fp}. The resulting gamma-ray flux at Earth is given by
\begin{eqnarray}
\frac{d\Phi_\gamma^{\rm EG}}{dE_\gamma} &=& 
\frac{\Gamma_\chi(\gamma)}{ 4\pi M_\chi} 
\frac{f_\chi \rho_{\rm DM} c}{H_0} \\
&&\hskip -0.3cm\times 
 \int_0^{z_{\max}} \frac{dz}{1+z}   
\frac{  e^{- \Gamma_\chi t(z)}}{\sqrt{\Omega_m (1+z)^3 + \Omega_\Lambda}} P \Bigl (z, \frac{dN_\gamma^{\rm inj}}{dE_\gamma} \Bigr )  
\nonumber\label{eq:Total_gamma_flux} \quad \quad
\end{eqnarray}
where $\Gamma_\chi (\gamma)$ is the decay rates for relevant decay process contributing to photon production and $dN_\gamma^{\rm obs}/dE_\gamma \equiv P(z_0,dN_\gamma^{\rm inj}/dE_\gamma)$ is the operation including the attenuation of high-energy gamma-rays through pair-production ($\gamma + \gamma_{\rm bkd} \to e^+ + e^-$) and reproduction of down-scattered photons by inverse Compton scattering ($e^\pm + \gamma_{\rm bkd} \to e^\pm + \gamma$) in the propagation from the injection position at the redshift $z=z_0$ to the Earth ($z=0$). The cascade evolution can be checked by solving transport equation in the presence of CMB for microwave and Extragalactic Background Light (EBL) for the infra-red and optical components for each redshift value $z$. We adopt the EBL density at various redshift $z$ from Ref.~\cite{Saldana-Lopez:2020qzx} for our calculation. In this work, the evaluation has been done by utilizing GammaCascade \cite{Blanco:2018bbf, Capanema:2024nwe} which solve the transport equation semi-analytically. Similar to neutrino case, the injection spectrum $dN_\gamma^{\rm inj}/dE_\gamma$ was obtained by using HDMSpectra.

{\bf \textit{Signatures and observational constraints}}. Assuming $\tau_\chi = 1$ Gyr and $f_\chi = 0.6$, we consider one representative set of parameters for each scenarios as the following:
\begin{itemize}
\item Scenario I ($\chi_h \to \chi_l + \nu + h$)
\begin{eqnarray}
M_\chi & = & 10^{16} \text{ GeV}, \nonumber \\
\Delta M_\chi & = & 3 \times 10^9 \text{ GeV}, \nonumber \\  
\Lambda & = & 1.44 \times 10^{31} \text{ GeV}
\end{eqnarray}
\item Scenario II ($\chi_h \to \chi_l + \nu_s$)
\begin{eqnarray}
M_\chi & = & 6 \times 10^{12} \text{ GeV}, \nonumber \\
\Delta M_\chi & = & 5 \times 10^8 \text{ GeV}, \nonumber \\ 
\lambda_s & = & 1.12 \times 10^{-22}
\end{eqnarray}
\end{itemize}
which correspond to $\text{Br}_\nu \simeq 10^{-2.6}$ for Scenario I, and $\text{Br}_\nu \simeq 10^{-4} - 10^{-3}$ for Scenario II. 

 For Scenario I, although the scale $\Lambda$ is much greater than the Planck scale, it can be induced by sub-Planckian physics with technically natural small couplings. An example could include a heavy right-handed neutrino and its superpartner with mass around $m_N$, having Yukawa interactions with a Dirac neutrino yukawa $y_\nu$ and an additional yukawa coupling $y_N$ between heavy neutrino multiplet $N$ and chiral multiplet $X$. The superpotential is written as
\begin{eqnarray}
W & = & y_\nu L H_u N + y_N N X^2.
\end{eqnarray}
In this case the scale $\Lambda = m_N/(y_\nu y_N)$ appears from integrating out heavy field $N$. For instance, $y_\nu, y_N \sim 10^{-8}$ and $m_N \sim 10^{16}$ GeV can be a viable parameter choice in this scenario.

For this set of example parameters, our results are summarized in Figure.~\ref{fig:flux_EGdominant_DM_total}. In Scenario II, we set our parameters $\text{Br}_\nu \lsim 10^{-3}$ to satisfy the constraints on sterile-active mixing at $\mathcal{O}$(eV) mass \cite{Bolton:2019pcu}, and $\Delta M_\chi/M_\chi\sim 10^{-4}$ is taken to be consistent with the Lyman-alpha forest constraints on the kick velocity for the daughter dark matter particle~\cite{Wang:2013rha,Chou:2003wx}. 

We show the expected signal of neutrinos and gamma-rays and compare them with existing isotropic diffuse neutrino/gamma-ray flux limits. One finds the relevant constraints on each energy region of the spectrum in the caption of the figure.

\section{Conclusion and Discussion \label{sec:conclusion}}

In this letter, we have explored the Superheavy Supersymmetric Dark Matter (SSDM) scenario and its neutrino and gamma-ray signatures over a wide energy range. In the case of superheavy dark matter, supersymmetry naturally leads to a nearly degenerate mass spectrum among supermultiplet fields. The energy of neutrinos injected from the decay of heavier components carries direct information about the scale of SUSY breaking.

If neutrinos are produced via the direct decay of a heavier component, the process inevitably involves the production of Higgs bosons, leading to an additional gamma-ray signal. In contrast, if neutrinos originate from the oscillation of sterile neutrinos produced in the decay of a heavy component, the resulting neutrino spectrum is expected to be sharper, with no accompanying gamma-ray signal.

Motivated by the recent detection of an ultra-high-energy (UHE) neutrino event at $E_\nu \sim 220$ PeV by the KM3NeT collaboration, we have analyzed this scenario in the context of SSDM. Our findings indicate that if the lifetime of the heavier component of dark matter is around $\mathcal{O}$(1) Gyr, an extragalactic, nearly isotropic neutrino flux is robustly expected at energies around $E_\nu = \mathcal{O}(0.1-1)\Delta M_\chi$. Moreover, it is straightforward to consider scenarios in which the lifetime of heavier components is shorter, which could lead to distinctive observational consequences.

This scenario could have broader implications for the spectra of high-energy cosmic particles across various energy ranges, including ultra-high-energy cosmic rays (UHECRs) at or above EeV scales and very-high-energy gamma rays in the TeV-PeV range. With upcoming neutrino and ultra-high-energy cosmic ray observatories, such as IceCube-Gen2 \cite{IceCube-Gen2:2020qha} and AuperPrime \cite{Castellina:2019irv}, we anticipate that the nature of superheavy dark matter and the origins of ultra-high-energy cosmic particles will be further unveiled in the near future ~\cite{Jho:2018dvt, Jho:2020sku, Jho:2021rmn}.


\section*{Acknowledgments}
CSS thanks Eung Jin Chun and Seodong Shin for useful discussions about decaying dark matter. SCP thanks Carsten Rott and Kazunori Kohri.  
This work was supported by the National Research Foundation of Korea (NRF) grant funded by the Korea government (MSIT) RS-2022-NR072128 (CSS), RS-2023-00283129 (SCP) and RS-2024-00340153 (SCP, YSJ). SCP is also supported by Yonsei internal grant for Mega-science (2024-22-0590).)

\appendix
\section{R-parity violation, BBN safety, and proton stability}
\label{app:RPV}
In our supersymmetric superheavy dark-matter scenario, R-parity violation (RPV) is introduced to ensure that the MSSM LSP decays before BBN, while not endangering proton stability:
\begin{itemize}
\item The LSP decays promptly via RPV.
\item We allow either $B$ or $L$ violation, but not both, thereby removing the dangerous dimension-4 $p$-decay operators. A heavy gravitino kinematically forbids exotic channels such as $p\to K^+ + \psi_{3/2}$.
\end{itemize}
In this appendix, we qualitatively show how these conditions can be satisfied.

\subsection{BBN safety}
In high-scale SUSY the LSP is typically heavy; even if a relatively light neutralino arises from a small $\mu$-term, RPV-induced decays are fast enough provided the couplings are not extremely tiny.
For trilinear RPV with an off-shell sfermion mediator, the dominant three-body width follows the standard estimate \cite{Barbier:2004ez,Dreiner:1997uz}:
\begin{align}
\Gamma_{\rm 3body}(\mathrm{LSP}\!\to\! f f f)
&\sim
\frac{|\lambda_{\rm RPV}|^{2}}{192\,\pi^{3}}\,
\frac{m_{\rm LSP}^{5}}{\tilde m^{4}},
\label{eq:rpv-3body}
\end{align}
where $\lambda_{\rm RPV}\in\{\lambda,\lambda',\lambda''\}$ denote the RPV couplings associated with the superpotential terms $LLE$, $LQD$, and $UDD$, respectively; $m_{\rm LSP}$ is the LSP mass, and $\tilde m$ is a typical sfermion mass.
Numerically,
\begin{align}
\tau_{\rm LSP}
&\approx
3.9\times 10^{-14}\,\sec \nonumber\\
&\quad \times \left(\frac{10^{-7}}{\lambda_{\rm RPV}}\right)^{2}
\left(\frac{\tilde m}{10^{8}\,\mathrm{GeV}}\right)^{4}
\left(\frac{10^{5}\,\mathrm{GeV}}{m_{\rm LSP}}\right)^{5}.
\label{eq:rpv-numeric}
\end{align}
Thus, for a heavy LSP (e.g.\ $m_{\rm LSP}\!\gtrsim\!10^{5}\,\mathrm{GeV}$), the lifetime is far below $1$ sec even for very small $\lambda_{\rm RPV}$, ensuring BBN safety.
If a lighter neutralino appears (e.g.\ $m_{\rm LSP}\!\sim\!\mathrm{TeV}$), Eq.~\eqref{eq:rpv-3body} implies that a moderately small $\lambda_{\rm RPV}$ still suffices for $\tau_{\rm LSP} < 1\,\sec$ when $\tilde m$ is large. Hence, across the relevant mass ranges, the MSSM LSP is not cosmologically long-lived in our setup.

\subsection{Proton stability}
Dimension-4 proton decay requires the simultaneous presence of $B$- and $L$-violating couplings \cite{Hall:1984id,Barbier:2004ez}.
We therefore adopt one of the two safe choices

(i) $B$-conserving, $L$-violating RPV: $\lambda,\,\lambda'\!\neq\!0$, $\lambda''\!=\!0$

(ii) $L$-conserving, $B$-violating RPV: $\lambda''\!\neq\!0$, $\lambda,\,\lambda'\!=\!0$
Either choice eliminates the dangerous tree-level operators for $p\to \text{meson}+\ell$.
In the first case ($B$ conserved) standard proton-decay modes are absent; in the second case ($L$ conserved) they are also absent, though one must consider dinucleon decay and $n$–$\bar n$ transitions. These are parametrically suppressed here by the large sfermion masses and remain compatible with bounds for the small $\lambda''$ needed only to ensure pre-BBN LSP decay (see, e.g., \cite{Goity:1994dq}); discrete symmetries such as baryon triality $\mathbb{Z}_3^{B}$ or lepton parity can enforce these patterns and forbid mixed $B$- and $L$-violating operators \cite{Dreiner:2006zd}.
In our high-scale scenario the gravitino is very heavy ($m_{3/2}\!\gg\!m_p$), so proton-decay channels emitting a gravitino are kinematically forbidden.
Thus neither RPV nor supergravity-induced operators open viable $p$-decay modes in our setup.

Consequently, our SUSY implementation is compatible with current proton-decay limits while meeting the cosmological requirement that all MSSM relics decay well before BBN.
The only stable relic is the dark-sector state protected by its own $Z_2$, which constitutes today’s dark matter in our analysis.

\bigskip
\bibliographystyle{apsrev4-2}
\bibliography{ref}

\end{document}